\documentclass[aps,prb,twocolumn,groupedaddress]{revtex4-1}

\usepackage{graphicx}
\usepackage{dcolumn}
\usepackage{amsmath}
\usepackage[normalem]{ulem}

\makeatletter
\def\btt#1{\texttt{\@backslashchar#1}}
\DeclareRobustCommand\bblash{\btt{\@backslashchar}}
\makeatother

\usepackage{color}

\begin{document}

\preprint{HEP/123-qed}



\title[Short Title]{Electrically tunable three-dimensional $g$-factor anisotropy\\
in single InAs self-assembled quantum dots}

\author{
S. Takahashi$^1$, R. S. Deacon$^{2, 3}$, A. Oiwa$^{3, 4}$, K. Shibata$^{1, 5}$, K. Hirakawa$^{1, 3, 5}$, and S. Tarucha$^{1, 2, 4}$
}

\affiliation{
$^1$Institute of Nano Quantum Information Electronics, University of Tokyo, 4-6-1 Komaba, Meguro-ku, Tokyo 153-8505, Japan\\
$^2$Advanced Device Laboratory, RIKEN, 2-1 Hirosawa, Wako-shi, Saitama 351-0198, Japan\\
$^3$JST CREST, 4-1-8 Hon-cho, Kawaguchi-shi, Saitama 332-0012, Japan\\
$^4$Department of Applied Physics and QPEC, University of Tokyo, 7-3-1 Hongo, Bunkyo-ku, Tokyo 113-8656, Japan\\
$^5$Institute of Industrial Science, University of Tokyo, 4-6-1 Komaba, Meguro-ku, Tokyo 153-8505, Japan
}


\date{\today}


\begin{abstract}
Three-dimensional anisotropy of the Land\'{e} $g$-factor and its electrical modulation are studied for single uncapped InAs self-assembled quantum dots (QDs).
The $g$-factor is evaluated from measurement of inelastic cotunneling via Zeeman substates in the QD for various magnetic field directions.
We find that the value and anisotropy of the $g$-factor depends on the type of orbital state which arises from the three-dimensional confinement anisotropy of the QD potential.
Furthermore, the $g$-factor and its anisotropy are electrically tuned by a side-gate which modulates the confining potential.
\end{abstract}


\pacs{Valid PACS appear here}

\maketitle


The Land\'{e} $g$-factor, the magnetic response of spin, is a physical constant reflecting the spin-orbit interaction (SOI) and the quantum confinement effect in semiconductor nanostructures, since it is determined by coupling between orbital and spin angular momentum\cite{book:Winkler}.
In low-dimensional systems, the confining potential has large asymmetry resulting in an anisotropic $g$-factor.
Therefore, the $g$-factor anisotropy can be electrically modulated by gating the quantum dot (QD) confining potential.
This may be exploited for coherent manipulation of electron spins through $g$-tensor modulation resonance ($g$-TMR) which was previously studied for a quantum well\cite{paper:Kato}.
For single electron spins, self-assembled QDs (SAQDs) and nanowire quantum dots (NWQDs) made out of narrow gap semiconductors such as InAs, InP and InSb are relevant for the study of $g$-factor anisotropy\cite{paper:Mayer-Alegre, paper:Nakaoka, paper:Schroer, paper:Witek}, because they have quite large negative values of $g$-factor for electrons due to the strong SOI.
In particular for InAs QDs we previously demonstrated that both the SOI effect and the orbital states are influenced by three-dimensional (3D) electrostatic potential\cite{paper:Takahashi, paper:Kanai_1}.


Among SAQD systems InAs SAQDs are the most extensively studied in crystal growth as well as optical and electrical characterization.
The InAs SAQDs in our study are uncapped or unstrained so that the QD size is relatively large, laterally 100 nm wide and vertically 30 nm high.
The QD shape is anisotropic with the confinement strong in the out-of-plane direction and weak in the in-plane direction, leading to $g$-factor anisotropy\cite{paper:Takahashi}.
Furthermore, the in-plane confinement is so weak that the confinement potential and the confined electron wavefunction can be modulated by means of electrical gating.
A local or anisotropic gating has been applied to InAs SAQDs\cite{paper:Kanai_1, paper:Kanai_2, paper:Deacon} and NWQDs\cite{paper:Nadj-Perge} to modulate $in$-$situ$ QD-lead tunneling coupling, and angular anisotropy of SOI energy and $g$-factor.
In our previous study\cite{paper:Deacon} of the electrical tuning of $g$-factor we identified tunability of the $g$-tensor only in a two-dimensional (2D) plane and assumed that the QD could be approximated as a disk-like 2D harmonic potential as if often done\cite{paper:Jung, paper:Igarashi}.
Studies of the anisotropy of the SOI have however shown that a 3D confinement, arising from the QD shape as well as the metal electrodes asymmetrically contacted to the QD, may be more realistic\cite{paper:Kanai_1}.
Such a 3D confinement leads to arbitral direction of orbital angular momentum (OAM) and therefore arbitral $g$-factor anisotropy\cite{paper:Pryor} which can be modulated largely by electrical gating.


In this work, we investigated full 3D $g$-factor anisotropy for single InAs SAQDs.
The $g$-factor is derived by measurement of inelastic cotunneling or Kondo effect through the Zeeman substates in the QD\cite{paper:Sasaki, paper:Csonka}, since $g$-factor cannot be evaluated adequately via magnetic evolution of ground states due to the large electron charging energy\cite{paper:Andergassen}.
The obtained $g$-factor shows a different 3D anisotropy depending on the charge state.
From this behavior we distinguish the related orbital type as either that with small OAM type ($s$-orbital like) or large OAM ($p$-orbital like) confined to a 3D potential rather than a 2D potential.
We used an electrical sidegate (SG) placed nearby the QD to largely modulate the 3D QD confining potential and therefore the 3D $g$-factor anisotropy.
We finally characterize the performance of $g$-TMR in the present InAs SAQD using the values of $g$-factor anisotropy obtained here.


Uncapped InAs SAQDs were grown with Stranski-Krastanov mode by molecular beam epitaxy on a [001] semi-insulating GaAs substrate.
A pair of Ti (3 nm)/Al (150 nm) source and drain electrodes separated by a 30 nm gap was deposited on the surface of a single SAQD using electron beam lithography techniques\cite{paper:Jung}.
The QD is positioned at the edge of the nanogap and has a larger overlap with the drain lead than the source as shown in Fig.1(a).
Differential conductance of electron transport through the QD was measured for a small source-drain voltage $V_{sd}$ at a bath temperature of 40 mK using conventional lock-in techniques.
To change the number of electrons in the QD the backgate (BG) voltage $V_{bg}$ applied to a doped layer beneath the QD was modulated.
The $g$-factor was evaluated from the transport measurement for various magnetic field angles selected $in$-$situ$ using a vector magnet system.


\begin{figure}[t]
\includegraphics[width=0.9\linewidth]{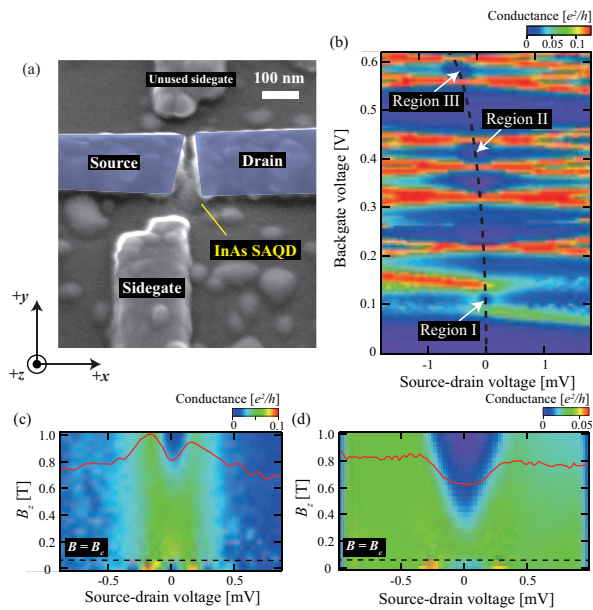}
\caption{\label{fig:sample}
(a) Scanning electron micrograph image of the device measured.
An InAs QD is contacted to the source and drain electrodes at the edge.
Only the sidegate nearest the QD is used for the present study.
The coordinates are defined as the figure.
(b) Differential conductance as a function of $V_{sd}$ and $V_{bg}$ measured for $B_z = 1$ T and $V_{sg}$ = 0 V.
The dashed line indicates the center of the Coulomb diamonds which is shifted due to leakage of the back-gate to the source lead as discussed in the main text.
(c) Magnetic evolution of inelastic cotunneling peaks and the Kondo anomaly in Region I.
Below 100 mT, superconducting features appear because the Al electrodes are superconducting.
The Kondo zero-bias anomaly is only observed when $|g_z|$$\mu_B$$B_z$ $<$ $k_B$$T_K$.
(d) Magnetic evolution of inelastic cotunneling steps in Region II.
The red traces in (c) and (d) show the data at $B_z$ = 0.8 T.
}
\end{figure}


Figure 1(b) shows the stability diagram with a perpendicular magnetic field $B_z$ = 1 T and a SG voltage $V_{sg}$ = 0 V.
We observe a series of diamond shaped regions or Coulomb blockade regions.
Note that the QD of interest is coupled with another small QD which exibits very small tunnel currents and has little effect on the transport discussed here\cite{EPAPS}.
The center of the diamond is shifted towards negative $V_{sd}$ as the BG voltage $V_{bg}$ is increased.
This shift is caused by current leakage from the BG to the source electrode and does not influence the transport characteristics of the QD, so we regard the dashed line as $V_{sd}$ = 0 V hereafter.
In three different Coulomb blockade regions, I, II, and III in Fig.1(b), two peaks for I and two steps for II and III are observed on both sides to the dashed line.
These peaks or steps are due to inelastic cotunneling through the QD, which appears when $eV_{sd}$ is consistent with excitation energy in the QD\cite{paper:Franceschi}.
The onset of inelastic cotunneling is usually signified by steps rather than peaks.
The peaks are only observed when the cotunneling is associated with the Kondo process or when the peak separation is not very large as compared to the Kondo temperature $T_K$\cite{paper:Sasaki}.
This is indeed the case for the present QD.


We measure the magnetic field evolution of the inelastic cotunneling peaks to study the inelastic cotunneling and the Kondo effect for the three regions.
The results measured for Region I and II are shown in Fig.1(c) and (d), respectively.
The Al electrodes have a superconducting critical field $B_c$ $\sim$ 100 mT and here we focus on the field range above 100 mT where the leads are in the normal state.
Note that two peaks observed at $V_{sd}$ $\sim$ $\pm$ 0.2 mV below 100 mT are caused by quasi-particle tunneling\cite{paper:Buizert, paper:Kanai_2} between the superconducting leads.


In Fig.1(c), the two inelastic cotunneling peaks merge to a zero-bias peak or Kondo peak as the magnetic field is decreased from $B_z$ = 1 T down to 500 mT.
The zero-bias peak is unchanged with $B_z$ down to 100 mT.
From the peak width we evaluate $T_K$ $\sim$ 1 K for Region I.
The two inelastic cotunneling peaks become separated approximately linearly with $B_z$, reflecting Zeeman energy $|g_z|\mu_BB_z$, where $\mu_B$ is the Bohr magneton.
From this behavior we evaluate the $g$-factor of $|g_z|$ = 3.5 for Region I.
The Zeeman energy is 0.1 meV at $B_z$ = 500 mT, comparable to $k_BT_K$, where $k_B$ is the Boltzmann constant.
Note that the Zeeman splitting appears only when the electron number in the QD is odd, and therefore the SU(2) Kondo effect appears at low magnetic fields.
For Region II the two inelastic cotunneling steps split by the Zeeman effect are observed at $|eV_{sd}|$ = $|g_z|$$\mu_{B}$$B_z$ with $|g_{z}|$ = 5.2 in Fig.1(d).
For Region III, similar inelastic cotunneling steps are observed but with a different $g$-factor of $|g_{z}|$ = 3.8 (not shown).
The Kondo zero-bias anomaly is not clear in Region II and III, probably because $T_K$ is lower than the measurement temperature.


\begin{figure}[t]
\includegraphics[width=0.8\linewidth]{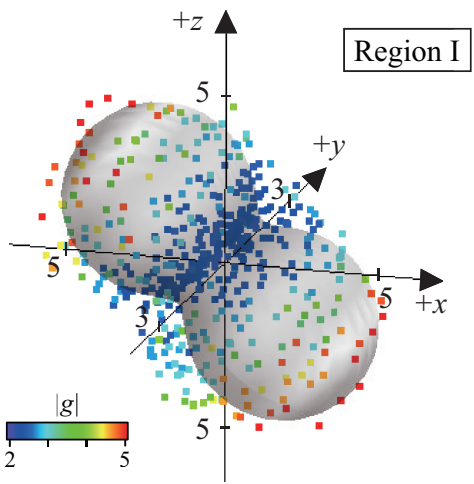}
\caption{\label{fig:3D}
3D polar plot of the evaluated $g$-factor (points) in Region I with $V_{sg}$ = 0 V for application of magnetic field in various directions and its 3D surface plot (gray) obtained by the theoretical fit to the data using the equation in the main text.
The distance of each plotted point from the origin is a measure of the absolute value of the $g$-factor which is also indicated by color.
}
\end{figure}


Then we set the magnetic field at $B$ = 1 T directed with various angles $\theta$ measured from the $+z$ axis in the $x$-$z$ and $y$-$z$ plane and $\phi$ measured from the $+x$ axis in the $x$-$y$ plane, and measured the inelastic cotunneling peaks (or steps) split by $\Delta$($\theta$, $\phi$) in $V_{sd}$.
We evaluated the $g$-factor at various sets of $\theta$ and $\phi$ as given by $e\Delta$/$\mu_BB$, and derived 3D magnetic angular dependence of the $g$-factor as shown in Fig.2 (points).
Color indicates the absolute value of the $g$-factor in each magnetic field direction represented as a distance from the coordinate origin.
In this figure the $g$-factor takes a maximum in the direction approximately $(\theta, \phi) = (-45^{\circ}, 180^{\circ})$ and $(135^{\circ}, 0^{\circ})$ tilted from the $z$-axis by $-45^{\circ}$ in the $x$-$z$ plane, and takes a minimum in the direction about $(45^{\circ}, 0^{\circ})$ and $(-135^{\circ}, 180^{\circ})$.


General model for a 3D anisotropic $g$-factor\cite{paper:Schroer} is
$g(B) = \sqrt{g_{1}^{2}B_{1}^{2}+g_{2}^{2}B_{2}^{2}+g_{3}^{2}B_{3}^{2}}/|B|$,
where ($g_1$, $g_2$, $g_3$) are the $|g|$ values in the orthogonal principal axes directions and ($B_{1}$, $B_{2}$, $B_{3}$) are the magnetic field components along the principal axes.
These may be defined as $B_{1} = B\cos(\phi+\phi_{0})\sin(\theta+\theta_{0})$, $B_{2} = B\sin(\phi+\phi_{0})\sin(\theta+\theta_{0})$ and $B_{3} = B\cos(\theta+\theta_{0})$.
Here $\phi_{0}$ and $\theta_{0}$ define the offset of the $g$-factor principal axes from the measurement axes.
Figure 2 also shows a 3D surface plot of the $g$-factor in Region I with $V_{sg}$ = 0 V fitted by the above equation with the following parameters, $(g_1, g_2, g_3, \theta_0, \phi_0) = (5.2, 2.3, 2.2, -49^{\circ}, 12^{\circ})$.


The $|g|$ values in the cut plane $x$-$y$, $y$-$z$, and $x$-$z$ in Fig.2 are shown in Figs.3(a), (b), and (c), respectively.
The $g$-factor is anisotropic with magnetic field directions in all figures and the anisotropy is significantly different from figure to figure.
In the $x$-$y$ cut plane of (a), the $g$-factor takes a maximum (minimum) of $|g_{max}|$ = 4.1 ($|g_{min}|$ = 2.4) at $\phi$ = $-8^{\circ}$ measured from the $x$-axis.
Similar shape of anisotropy is observed in the $y$-$z$ cut plane of (b) with the maximal $g$-factor at $\theta$ = $-11^{\circ}$ measured from the $z$-axis.
The $g$-factor anisotropy is the largest in the $x$-$z$ cut plane of (c) with $|g_{max}|$ = 4.9 and $|g_{min}|$ = 2.2 at $\theta$ $\sim$ $\mp$$45^{\circ}$ measured from the $z$-axis, respectively.
These results differ from our previous study for a QD with geometrically symmetric coupling to the leads\cite{paper:Takahashi} in which the $g$-factor was isotropic for in-plane rotation and maximal in the $z$-direction.
The $g$-factor in QDs strongly depends on the symmetry of the confinement which affects OAM\cite{paper:Pryor}.
The larger the OAM the larger the $|g|$ value.
Therefore, the OAM of electrons in Region I is the largest when the magnetic field is applied in the $x$-$z$ plane with $\theta$ $\sim$ $-45^{\circ}$.
The same measurement and theoretical fitting of the $g$-factor anisotropy as shown in Fig.2 were performed for Region II and III.
The results for the $x$-$z$ cut plane are shown in Fig.4(a), and (b) for Region II, and III, respectively.
The $g$-factor anisotropy in Region II is similar to that in Region I, showing a maximum at $\theta$ $\sim$ $-45^{\circ}$ with $|g_{max}|$ = 5.4.
For Region III, however, the $g$-factor is almost isotropic with $|g|$ = 4, which is the same for the other planes, indicating a spherical distribution of the $g$-factor.
This difference may be accounted for by assuming that the orbital type is $p$-like (large OAM) for Region I and II, whereas it is $s$-like (small OAM) for Region III.


\begin{figure}[t]
\includegraphics[width=1.0\linewidth]{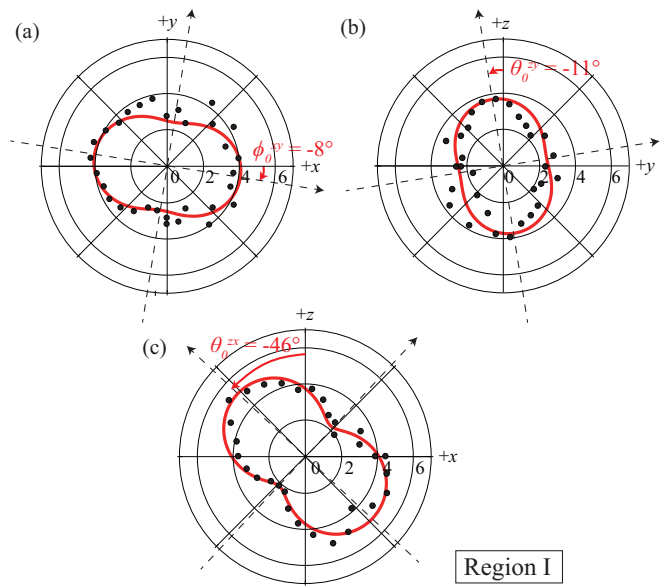}
\caption{\label{fig:projections}
The $g$-factor in the cut plane of $x$-$y$ (a), $y$-$z$ (b), and $x$-$z$ (c) of the 3D polar plot of data in Region I at $V_{sg}$ = 0 V presented in Fig.2.
The red line is the theoretical fit.
The dotted lines indicate the axes of the maximum and minimum $g$-factor directions.
}
\end{figure}


We assume that the tilting of the $g$-factor principal axis from the $z$-axis, observed for Region I and II, is due to the 3D asymmetry of the QD confining potential.
In Fig.1(a) the QD is positioned at the edge of the nanogap and approximately one half of the QD is covered by the drain electrode metal while a small part of the QD is covered by the source electrode metal.
In our previous work\cite{paper:Takahashi}, we discussed that electrons in the QD beneath the source and drain electrode metal are depleted.
Therefore, the potential shape of the QD is considered as a half pyramid in the uncovered region by the contact electrodes\cite{EPAPS}.
We used an 8-band {\bf k}$\cdot${\bf p} theory to calculate the probability density of electrons confined by such an exotic potential\cite{nextnano}.
A $p$-orbital state confined by a 2D potential extends over the $x$-$y$ plane, having the largest magnetic coupling with a magnetic field along the $z$-axis.
In the present case, however, a $p$-like orbital state extends over a plane tilted by about $+45^{\circ}$ from the $z$-axis\cite{EPAPS} and therefore the $g$-factor principal axis is tilted by $-45^{\circ}$ from the $z$-axis\cite{paper:Pryor}, which is consistent with our finding in Fig.3(c) and Fig.4(a).
To further examine the 3D confinement, we studied the in-plane magnetic field effect on the Coulomb peaks\cite{EPAPS}.
The in-plane magnetic field evolution of states is not linear, indicating that not only Zeeman effect but also OAM effect are involved and that OAM may not be perpendicular to the in-plane magnetic field in the 3D confinement potential.
Note that in another sample having a QD symmetrically covered by the source and drain electrode metal, the observed $g$-factor has a principal axis along the $z$-axis and it is almost isotropic with respect to the in-plane magnetic field directions\cite{EPAPS}.


\begin{figure}[t]
\includegraphics[width=1.0\linewidth]{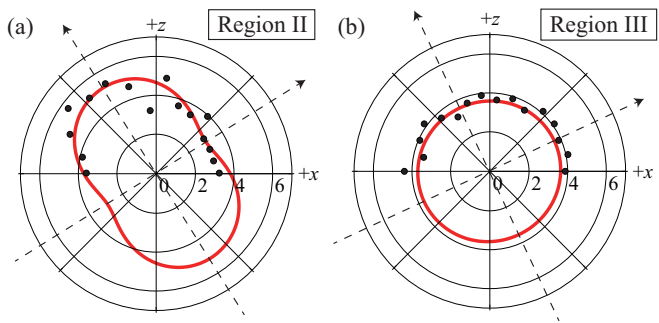}
\caption{\label{fig:different regions}
Anisotropic $g$-factor for Region II(a) and III(b) with $V_{sg}$ = 0 V in the $x$-$z$ plane plotted in the same way as shown in Fig.3(c).
}
\end{figure}


\begin{figure}[t]
\includegraphics[width=0.8\linewidth]{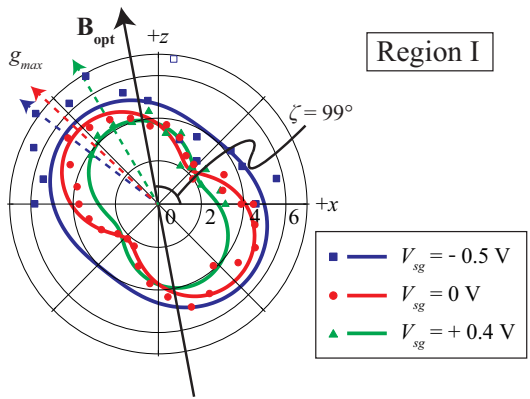}
\caption{\label{fig:sidegate}
Anisotropic $g$-factor in the $x$-$z$ plane measured for Region I at $V_{sg}$ = -0.5 V in blue, 0 V in red, and +0.4 V in green.
As $V_{sg}$ is made positive, the $g$-factor get small and the $g$-factor maximal direction $g_{max}$ approaches the $z$-axis.
The optimized direction of magnetic field $\bf B_{opt}$ is selected as $\Omega_{\perp}$ becomes the largest at $V_{sg}$ = 0 V (explained in the text).
}
\end{figure}


We now discuss use of the SG to modify the $g$-factor anisotropy.
The electric line of force due to the SG voltage $V_{sg}$ extends through the GaAs substrate with its large permittivity, so that application of positive $V_{sg}$ may shift the electron wavefunction downwards in the direction of the SG.
Figure 5 shows the $g$-factor anisotropy measured for magnetic field in the $x$-$z$ plane at various $V_{sg}$ of -0.5, 0, and +0.4 V in Region I.
As $V_{sg}$ is made positive, the amplitude of the $g$-factor becomes small, and the $g$-factor maximal direction $g_{max}$ approaches the $z$-axis.
This behavior is qualitatively predicted from our simulation\cite{EPAPS} which shows that the OAM vector of the $p$-like orbital state tends to be tilted toward the $z$-axis so does the $g_{max}$ direction as $V_{sg}$ is made positive.
When the electron wavefunction shifts to the bottom of the QD the $|g|$ value may become small, because Ga atoms are diffused from the GaAs substrate into the QD near the substrate\cite{paper:Shibata, paper:Kegel}.


Finally, we estimate the possible Rabi frequency of electron spin by means of $g$-TMR from the obtained results.
The anisotropic $g$-factor is described by a tensor, $g_{ij}$ ($i,j$ = $x, y, z$).
For application of a static magnetic field $\bf{B_0}$, the spin precession vector $\bf{\Omega_0}$ is given by $\Omega_i$ = $\sum_{j}g_{ij}B_j$.
Then suppose to apply microwave (MW) on the SG.
Because the $g$-factor anisotropy or $g$-tensor depends on $V_{sg}$ as shown in Fig.5, the MW induced a.c. voltage gives rise to an a.c. precession vector which has a component $\bf{\Omega}_{\perp}$ perpendicular to the initial $\bf{\Omega_0}$.
The $\bf{\Omega}_{\perp}$ is maximal for $\bf{B_0}$ $\parallel$ $\bf{B}_{opt}$ in Fig.5.
For example, for the Larmor precession frequency of 20 GHz with $\bf{\Omega}$ and MW induce a.c. voltage of 1 mV, the calculated Rabi frequency is 0.88 MHz.
This value is similar to that in our previous work\cite{paper:Deacon}.


In summary, we have investigated 3D $g$-factor anisotropy in single InAs SAQDs by measuring inelastic cotunneling for three different charge states.
We observed that the $g$-factor principal axis is tilted by about $45^{\circ}$ from the growth direction, and assigned it to the confinement asymmetry created by the QD shape as well as the geometry of the source and drain electrodes.
We observed anisotropic and isotropic $g$-factor depending on the charge state related to the orbital type.
In addition, we succeeded in electrically tuning the $g$-factor anisotropy via the confinement potential modulation with $V_{sg}$.
The $g$-factor anisotropy can be optimized toward a manipulation of electron spins by means of $g$-TMR in the QDs by appropriately designing the QD shape, confinement potential, and the coupling efficiency to the SG.


S. Takahashi gratefully thanks Y. Arakawa for encouragement.
Part of this work was supported by Grant-in-Aid for Research A (No. 21244046), and Innovative Areas (21102003) MEXT, JST strategic Int. Coop. Program, FIRST program, IARPA 'Multi-Qubit Coherent Operations' through Copenhagen University, DARPA QuEST grant (HR-001-09-1-0007), and MEXT Project for Developing Innovation Systems.



\end{document}